• Article •

# Exploring the Reform and Development Pathways of AIIB's Climate Accountability Mechanism in the Context of Global Climate Governance

**Yedong Zhang**[1],*

[1] School of Law, Fudan University, Shanghai, China.

* Corresponding Authors: Yedong Zhang. Email: zyedong1422@163.com

**Abstract:** After the Asian Infrastructure Investment Bank (AIIB) revised its Environmental and Social Framework, it has committed to certain climate-related objectives, yet an independent climate accountability mechanism has not been established. The absence of clear evaluation principles and procedural rules presents challenges in effectively addressing environmental investment disputes. This article reviews both domestic and international literature related to AIIB's climate accountability mechanism, identifying that the current Environmental and Social Framework's principled content and reliance on traditional approaches have resulted in issues of enforceability and the absence of an independent accountability institution. To enhance the effectiveness of AIIB's climate accountability mechanism, a reassessment of its development path is necessary. Future developments should transition from an additive path to a substitutive path, focusing on the application of international environmental and social standards, increasing stakeholder recognition and participation, promoting comprehensive reforms within AIIB, and establishing a coordinated independent accountability system. These measures aim to support the robust development of AIIB's climate accountability mechanism.

## 1. Introduction

In the face of escalating global climate challenges, the Asian Infrastructure Investment Bank (AIIB) has emerged as a pivotal player in fostering sustainable development through its Climate Accountability Mechanism. This paper explores the reform and development pathways of AIIB's Climate Accountability Mechanism, emphasizing its integral role in the global quest for effective climate governance. The climate accountability mechanism is an indispensable part of modern international financial institutions in promoting sustainable development and environmental protection. Its core lies in ensuring that loans and investment projects fulfill their environmental and social responsibilities, thereby avoiding negative impacts on local ecosystems and communities. For the Asian Infrastructure Investment Bank (AIIB), strengthening the climate accountability mechanism is not only a pressing response to global climate change but also a crucial aspect of enhancing its international credibility and commitment to sustainable development. Traditional international financial institutions like the World Bank began implementing stringent environmental and social policies decades ago, epitomized by the introduction of the Equator Principles. These principles require signatory banks to consider environmental and social impacts when assessing and managing project risks. The widespread application of these principles has set a benchmark for responsible investment by international financial institutions. The Equator Principles serve as a risk management framework adopted by financial institutions for determining, assessing, and managing environmental and social risk in projects. They are primarily intended to provide a minimum standard for due diligence to support responsible risk decision-making. As a relatively new international financial institution, AIIB can avoid potential environmental and social risks and enhance its competitiveness and influence in the international financial market by adhering to and promoting these high standards. This adherence not only mitigates risks but also aligns AIIB with global best practices, thus positioning it as a leader in responsible and sustainable financing. The AIIB's commitment to these principles underscores its dedication to mitigating the adverse effects of its investments and ensuring long-term benefits for the communities involved.In May 2021, the AIIB's Board of Directors reviewed and approved the revised Environmental and Social Framework, aiming to achieve 50% climate financing of approved financing by 2025, further strengthening its commitment to combating climate change. This ambitious goal signifies AIIB's proactive stance in addressing climate change and its recognition of the critical role that financial institutions play in promoting sustainability. The key focus of this policy framework is the construction, development, and improvement of the climate accountability mechanism. By establishing clear and enforceable standards, AIIB aims to ensure that all funded projects adhere to rigorous environmental and social guidelines. This framework is designed to provide a comprehensive approach to risk management, ensuring that projects do not contribute to environmental degradation or social inequity. In supporting infrastructure construction, international financial institutions' loan projects may negatively impact local living conditions. To avoid disputes and protect investment interests, international financial institutions such as the World Bank and AIIB generally follow the Equator Principles, setting high standards for environmental and social policies, ensuring that projects meeting these standards receive bank funding. The adoption of these principles ensures a level of consistency and accountability across projects, fostering a culture of sustainability and responsibility. Additionally, these institutions establish independent climate accountability mechanisms (e.g., the World Bank's Inspection Panel) to provide compensation to those affected by the projects. Such mechanisms are vital for addressing grievances and ensuring that affected communities have a voice in the development process. They provide a structured process for addressing complaints and ensuring that environmental and social standards are upheld. Therefore, the development and enhancement of AIIB's climate accountability mechanism are crucial for its success as a sustainable financial institution. By adopting and promoting high standards, strengthening its framework, and ensuring independent oversight, AIIB can lead the way in responsible investment and significantly contribute to global efforts in combating climate change and promoting sustainable development.

Reviewing the development history of the Asian Infrastructure Investment Bank (AIIB), significant efforts have been made in building its climate accountability mechanism. Since its establishment in 2015, AIIB has continuously improved its framework and policies concerning environmental and social responsibilities. From the initial draft of the Environmental and Social Framework in 5 to the official version in 2016, and further revisions in 2019 and 2021, AIIB has progressively established a systematic evaluation and management mechanism. These efforts are reflected not only in the issuance of policy documents but also in their strict implementation and supervision in practice. In 2015, as a new international financial institution in the Asian region, AIIB officially announced the draft Environmental and Social Framework. This framework aimed to assess potential environmental and social impacts according to relevant standards during its financing activities, thereby mitigating adverse effects on the environment and society of the project locations. In 2016, the official version of the Environmental and Social Framework was published, and in February 2019, AIIB made its first revision, followed by another revision in 2021 to further improve the institutional system. While AIIB has established a climate accountability mechanism, it continues to face practical challenges. Firstly, the issue of independence remains; the current mechanism lacks sufficient independence and transparency, which may allow for administrative interference. This situation complicates AIIB's ability to ensure that environmental and social standards are upheld impartially. For an accountability mechanism to be effective, it is essential that the institution operates without undue influence, ensuring that all assessments and decisions are based solely on established environmental and social criteria. Secondly, there are challenges related to enforcement; despite having a clear policy framework, the implementation of specific projects varies, with some projects lacking comprehensive environmental and social impact assessments. This inconsistency in project implementation affects the credibility of AIIB's commitment to environmental and social responsibility. To improve this, AIIB should ensure that all funded projects undergo rigorous and uniform evaluation processes, consistently applying environmental and social standards. Additionally, there is a need to strengthen public participation and transparency. Ensuring the full participation and right to information of affected groups is vital for achieving accountability. Effective public participation involves not only informing stakeholders about potential impacts but also actively involving them in decision-making processes. This approach enhances the legitimacy of AIIB's projects and ensures that the concerns and needs of affected communities are adequately addressed. Existing literature has made some review of AIIB's environmental accountability mechanisms, such as studies on complaint and accountability mechanisms for investment disputes involving environmental issues in AIIB, reflections on AIIB's environmental and social protection policies, and analysis of loan conditions, and the study of AIIB's operational system itself. However, they have not made systematic and targeted analyses of AIIB's climate accountability mechanism. Therefore, while AIIB has implemented a climate accountability mechanism, its effectiveness is influenced by challenges related to independence and enforcement. These challenges impact the mechanism's operational efficiency and overall effectiveness. In terms of improving the climate accountability mechanism, AIIB can draw on the successful experiences of other international financial institutions. For example, the World Bank's Inspection Panel has high standards of independence and transparency, ensuring the fairness and effectiveness of the accountability mechanism through publicly available investigation reports and feedback mechanisms from affected groups. AIIB can make improvements in these areas to enhance the authority and credibility of its climate accountability mechanism. Strengthening AIIB's climate accountability mechanism requires a multi-faceted approach. Firstly, AIIB needs to enhance the independence of its accountability framework by establishing independent oversight bodies that can operate without interference. These bodies should have the authority to conduct impartial assessments and enforce compliance with environmental and social standards. Secondly, AIIB should standardize its enforcement procedures to ensure consistency across all projects. This involves developing clear guidelines and rigorous evaluation criteria that are uniformly applied to all funded projects. Thirdly, increasing public participation and transparency is essential. AIIB should implement mechanisms that facilitate stakeholder engagement and ensure that affected communities have a voice in decision-making processes. This includes public consultations, transparent reporting, and accessible grievance mechanisms. By addressing these areas, AIIB can develop a more robust and effective climate accountability mechanism. This will not only enhance AIIB's credibility and operational effectiveness but also contribute significantly to global sustainable development efforts. The following sections will analyze the potential deficiencies and causes of AIIB's climate accountability mechanism based on existing domestic and international literature and propose a path for the healthy development of AIIB's climate accountability mechanism to improve it.

## 2. Evaluation of AIIB's Climate Accountability Mechanism: Identifying Challenges and Underlying Factors

As disputes from Belt and Road Initiative projects increase, effective resolution mechanisms are often lacking due to the absence of an AIIB-led dispute resolution mechanism and the independent operation of its accountability system. The challenges in AIIB's climate accountability mechanism can be attributed to the principled content of its Environmental and Social Framework, reliance on traditional methods, issues with enforceability, and the absence of an independent climate accountability institution. Both internal and external factors can impact the fairness and effectiveness of the mechanism. To address these challenges, AIIB should consider transitioning from additive to substitutive reform paths in developmental finance. This transition involves strengthening the application of international environmental and social standards, enhancing stakeholder participation, promoting institutional reforms, and establishing a coordinated independent accountability system. Strengthening international standards will ensure that projects are assessed and managed according to widely accepted environmental and social criteria, improving transparency and accountability. Enhancing stakeholder participation ensures that all relevant parties can voice their concerns and contribute to decision-making processes, leading to more inclusive and equitable outcomes. Promoting institutional reforms within AIIB will help balance power among member countries, reducing the disproportionate influence of developed countries and increasing input from developing countries. This balance is crucial for creating a more equitable framework that addresses the needs and concerns of all stakeholders. Establishing a coordinated independent accountability system is essential to ensure that AIIB's projects do not cause or contribute to environmental harm. This system would involve setting up independent bodies to oversee and enforce compliance with environmental and social standards, providing a reliable mechanism for addressing grievances and ensuring that affected communities receive adequate remedies. These efforts will help AIIB develop a robust climate accountability mechanism, improving its operational effectiveness and contributing to global sustainable development. By adopting these strategies, AIIB can enhance its role in the international financial landscape. This transformation will not only boost AIIB's credibility and effectiveness but also significantly contribute to global sustainable development. The causes of these challenges are analyzed as follows:

**2.1 Independence Considerations**

AIIB's climate accountability mechanism currently does not include a fully independent body for supervision and execution. This situation can lead to the susceptibility of the accountability mechanism to internal and external influences, which may impact its fairness and effectiveness despite the clear environmental and social standards outlined in the Environmental and Social Framework. Internal management decisions might adjust or overlook environmental and social assessment results due to economic interests and project advancement considerations. Additionally, external political and economic factors, such as member states' interests and international political pressures, can also affect the implementation of the accountability mechanism.AIIB currently lacks specific provisions for the climate accountability mechanism, particularly in areas such as due diligence, where national and corporate systems are sometimes substituted for bank policies, and in the establishment of grievance mechanisms. Chris Humphrey notes that while there is room for improvement in the existing systems of multilateral development banks, AIIB has met international expectations by referencing and adopting these established frameworks. As AIIB continues to develop a framework akin to other multilateral development banks, concerns regarding its multilateral nature are expected to diminish. However, being a newly established institution, AIIB has not yet fully developed its accountability mechanism, which currently lacks the capacity to handle large-scale investment disputes and complex international affairs. The absence of independent oversight presents challenges to AIIB's performance in environmental governance and sustainable development, highlighting the need for institutional reform and the establishment of independent bodies to enhance the effectiveness and fairness of its accountability mechanism.

## 2.2 Dependence on Established Approaches

As an emerging multilateral development bank, AIIB's institutional design and policy-making are significantly influenced by traditional multilateral development banks. This dependence on established approaches can pose challenges to innovation and adaptation within AIIB. While drawing on traditional experiences provides valuable references, it also limits AIIB's capacity to innovate and tailor its strategies to its specific needs, potentially resulting in an ineffective climate accountability mechanism.Multilateral development banks, including AIIB, encounter substantial challenges in institutional innovation. Despite AIIB's efforts to distinguish itself from traditional development banks and establish a new institutional framework, this reliance on established practices can constrain the negotiators' ability to innovate or may result in a lack of necessary resources and motivation for institutional innovation. Although AIIB aimed to develop a distinct environmental and social protection system, different from traditional multilateral development banks, to avoid broad assessment scopes, high project costs, and rigid mechanisms, its Environmental and Social Framework has gradually aligned with traditional practices after multiple revisions, reintroducing these issues.   Some scholars argue that this reliance on established approaches may lead AIIB to lower review standards by reducing environmental and social standards and might hinder the establishment of an independent accountability review mechanism. As a new type of multilateral development bank, AIIB should ensure that institutional innovation aligns with sustainable development and environmental and social governance principles. This reliance on traditional practices can limit the climate accountability mechanism's ability to adapt to AIIB's specific needs and conditions, reducing its flexibility and effectiveness in addressing contemporary climate challenges. While traditional models provide mature experiences, they may not fully reflect the current complexity and urgency of climate change. For example, in due diligence and risk assessment, AIIB continues to employ traditional methods, which may not effectively evaluate and respond to new risks and challenges posed by climate change. This conservative approach can reduce AIIB's flexibility and foresight in addressing climate change. To effectively tackle global climate change, AIIB must innovate beyond traditional practices, develop policies that reflect its unique characteristics, and design more flexible and comprehensive environmental and social risk assessment methods. These steps will enhance the overall effectiveness of its climate accountability mechanism.

## 2.3 Clarity and Operability of Provisions

The latest revision of the Environmental and Social Framework includes consultation procedures but lacks specific accountability provisions. Many of the provisions within the Environmental and Social Framework are broadly defined and lack specific implementation details and enforceability. AIIB faces resource and capacity constraints, particularly in developing its climate accountability mechanism, leading to a shortage of professional staff and technical support. This makes it challenging to meet the high standards expected in environmental and social assessments, risk management, and the design and implementation of accountability mechanisms.AIIB's relatively passive stance in cooperation with other international financial institutions also limits its capacity for institutional innovation and enhancement.

This issue is particularly evident in the standards for environmental and social impact assessments. Although the framework proposes high standards, the specific assessment methods and indicators remain unclear, leading to subjectivity and uncertainty in practice. This ambiguity makes it difficult to ensure consistency and fairness in project evaluations.The framework lacks clear evaluation principles and procedural rules for the legal application of complaint and accountability mechanisms, and it does not specify the bodies responsible for executing accountability, nor does it establish execution and supervision mechanisms. This can undermine the resolution of investment disputes involving environmental issues through privatized means. Moreover, the grievance and compensation mechanisms lack detailed procedures and standards, which can make it difficult for affected groups to seek fair treatment and reasonable compensation.European members of the World Bank played an influential role in the development of the Environmental and Social Framework, promoting the inclusion of global standards found in other multilateral development banks into AIIB. The framework, based on environmental and social policies (ESP), includes three sets of mandatory environmental and social standards (ESSs) for environmental and social assessment and management, involuntary resettlement, and indigenous peoples, along with an environmental and social exclusion list and guidelines for environmental and social policies. This provides AIIB with a comprehensive framework of principles for managing environmental and social risks and impacts. However, the accountability mechanism still lacks enforceability and binding power. The absence of clear grievance channels and standards complicates the process for affected groups to express their grievances and obtain appropriate compensation through formal channels. This ambiguity and lack of operability constrain the practical application of the Environmental and Social Framework in projects, potentially weakening its role in protecting environmental and social interests. Therefore, it is essential to refine the provisions and clarify the implementation details to enhance the framework's operability and enforceability, ensuring that affected groups receive fair and reasonable treatment.

## 3. Suggestions for the Healthy Development Path of AIIB's Climate Accountability Mechanism

From the development path of AIIB's climate accountability mechanism, it is evident that it is evolving from an additive path—the coexistence of new rules established by emerging powers alongside the old rules dominated by established powers—to a more advanced substitutive path, wherein new rules established by emerging powers replace the old rules. This transition marks a significant shift in the approach to international developmental finance. The low institutional flexibility of the World Bank as an international standard has gradually prompted China to adopt a substitutive reform path in developmental finance. Therefore, to promote the healthy development of AIIB's overall accountability mechanism, it is necessary not only to reform the superficial rule system but also to delve into deeper issues such as value orientations and the power configuration among member countries. Addressing these foundational elements is essential for creating a robust framework that supports sustainable development. Furthermore, it is crucial to discuss the construction and improvement of a coordinated independent accountability system through cooperative means. This involves fostering collaboration among member countries, enhancing transparency, and ensuring that the accountability mechanisms are both independent and effective.The specific aspects that need to be addressed include the following four points: strengthening the application of international environmental and social standards, increasing stakeholder recognition and participation, promoting comprehensive reforms within AIIB, and building a coordinated independent accountability system. By focusing on these areas, AIIB can ensure that its climate accountability mechanism evolves in a way that is not only more aligned with contemporary global standards but also more effective in promoting sustainable development and environmental protection.

### 3.1 Strengthening the Application of International Environmental and Social Standards

International financial institutions that hold significant positions in addressing environmental issues should rigorously strengthen the application of international human rights standards, and the Asian Infrastructure Investment Bank (AIIB) is a prime example of such an institution. Green development is its distinctive color and characteristic, highlighting its commitment to sustainable practices. The limitations imposed on loan obligations and additional conditions by AIIB will profoundly influence the plans and measures adopted by recipient countries as they advance their projects. This influence extends to whether timely, adequate, and effective remedies can be provided in the event of environmental damage. For AIIB itself, it is crucial at the institutional level to ensure that its projects do not cause, contribute to, or exacerbate environmental harm. This requirement serves as a necessary foundation for affected subjects to hold the institution accountable and to obtain appropriate remedies. By rigorously applying international environmental standards, the compliance review and supervisory evaluation of projects can become more normatively certain, thereby establishing a more robust climate accountability mechanism that effectively protects individual environmental rights. AIIB should prioritize enhancing the value of environmental protection and avoid partnerships with member countries that disregard these principles. Incorporating internationally recognized environmental standards into AIIB's policy framework can not only positively impact the fulfillment of environmental protection obligations by its member countries but also significantly enhance the legitimacy of the institution's actions and the credibility and effectiveness of its climate accountability mechanism. Strengthening the application of international environmental standards is fundamentally consistent with AIIB's overarching sustainable development goals and is essential for the healthy and effective development of its climate accountability mechanism. This comprehensive approach ensures that AIIB's initiatives are both ethically sound and practically effective in promoting global environmental stewardship.

### 3.2 Increasing Stakeholder Recognition and Participation

Ensuring that all relevant parties can participate equally in dispute resolution processes is crucial for achieving fair and just outcomes. Special attention should be paid to the protection of the procedural rights of complainants, as this is an essential prerequisite for genuine accountability. Due to the stringent restrictions of eligibility reviews, many complaints cannot advance to the substantive handling stage, and even fewer result in action plans that are fully implemented through the accountability mechanism. Given the inherent complexity of the complaint process as a prerequisite for accountability and the often unequal positions of the disputing parties, it is exceedingly necessary to provide comprehensive assistance to the complainants. In this regard, AIIB can take proactive measures by employing dedicated internal staff to provide direct assistance to complainants. Additionally, AIIB should encourage complainants to seek support from external experts, non-governmental organizations (NGOs), and other relevant entities. By increasing participation and offering robust support, the demands and concerns of stakeholders can be better expressed and understood, thereby facilitating the rational and fair resolution of issues. This approach not only enhances the transparency and credibility of the dispute resolution process but also ensures that all parties feel heard and respected, ultimately contributing to more equitable outcomes.

### 3.3 Promoting Overall Reform of AIIB

Promoting the overall reform of the Asian Infrastructure Investment Bank (AIIB) is highly beneficial for the development and enhancement of its accountability mechanism. When providing loans and technical assistance to developing and transitional countries, AIIB must ensure, at an institutional level, that developed countries cannot impose their preferences and interests on others through this mechanism. This approach should replace the current practice of addressing internal structural defects by overlooking environmental protection and shifting social responsibility. In the future, AIIB should strive to further refine the distribution of power within the organization and enhance the influence and discourse power of developing countries within its framework. This would significantly aid in the consistent application of stringent environmental standards and the overall improvement of the climate accountability mechanism. Such reforms would ensure that all member countries can coexist within a fair, reasonable, and orderly institutional system. Additionally, by addressing these structural imbalances, AIIB can remove unnecessary obstacles to the reform of its climate accountability mechanism, thereby fostering a more equitable and effective system for all stakeholders involved.

### 3.4 Building a Coordinated Independent Accountability System

To promote the effective implementation of AIIB's accountability mechanism, it is essential to establish a network of coordinated, open independent accountability mechanisms. These mechanisms must meet stringent requirements to ensure their effectiveness and credibility. Specifically, they should feature citizen-driven complaint and response systems, operate at the international level, be part of public institutions that fund or support development-related activities, maintain operational independence, and consider both social and environmental impacts or concerns comprehensively. The establishment of such a network of independent accountability mechanisms is crucial as it builds relatively stable and robust partnerships among institutions dedicated to accountability and redress. Through this network, institutions can collaborate on jointly funded projects, strengthen their interconnections, and actively exchange and discuss governance issues and best practices. This collaborative approach not only enhances their capabilities in accountability and compliance but also fosters a more transparent and responsible financial environment. By integrating these mechanisms, AIIB can ensure that its projects adhere to the highest standards of accountability, benefiting both the institution and the broader community.

## 4. Conclusion

In summary, as an emerging multilateral development bank, the Asian Infrastructure Investment Bank (AIIB) currently faces challenges in establishing a fully independent, comprehensive, and reliable climate accountability mechanism. The issues related to enforceability and the ambiguity of standards may impact the effective implementation of its environmental and social policies. A review of relevant domestic and international literature highlights that the challenges of AIIB's climate accountability mechanism stem from the broadly defined content of its Environmental and Social Framework, reliance on established approaches, and the absence of an independent climate accountability institution. To address these challenges, it is essential to reconsider the development path of AIIB's climate accountability mechanism. In the field of developmental finance, AIIB should gradually transition from an additive reform path—coexisting with traditional norms—to a substitutive reform path—establishing new standards. This transition requires systematic efforts to strengthen the application of international environmental and social standards, increase stakeholder recognition and participation, promote comprehensive institutional reforms, and build a coordinated, independent accountability system.By adopting these strategies, AIIB can enhance its role as a significant player in the international financial landscape and develop a distinctive and robust climate accountability mechanism. This transformation will enhance AIIB's credibility and effectiveness and significantly contribute to global sustainable development. Through these concerted efforts, AIIB can ensure that its initiatives are ethically sound, practically effective, and aligned with the highest standards of environmental and social responsibility. This holistic approach will ultimately position AIIB as a leader in promoting sustainable development and environmental stewardship on the global stage.

**Acknowledgement**

None.

**Funding Statement**

This research was funded by the Basic Research Funds for Universities in Inner Mongolia Autonomous Region (No. JY20220272), the Scientific Research Program of Higher Edu- cation in Inner Mongolia Autonomous Region (No. NJZZ23080), the Natural Science Foundation of Inner Mongolia (No. 2023LHMS05054), and the National Natural Science Foundation of China (No. 52176212).

**Author Contributions**

Jicai Guo: Writing, Original draft, Conceptualization, Methodology. Xiaowen Song: Conceptualization, Writing–review & editing, Supervision, Funding acquisition. Chang Liu: Data curation, Visualization. Yanfeng Zhang: Investigation, Formal analysis. Shijie Guo: Investi- gation, Data curation. Jianxin Wu: Supervision, Validation. Chang Cai: Supervision, Methodology. Qing'an Li: Project administration, Supervision, Funding acquisition. All authors reviewed the results and approved the final version of the manuscript.

**Availability of Data and Materials**

The data used in this study are confidential at the request of the wind farm operators.

**Conflicts of Interest**

The authors declare that they have no conflicts of interest to report regarding the present study.